# Thermal stabilization of thin gold nanowires by surfactant-coating: a molecular dynamics study

Stefan E. Huber[1,*], Chompunuch Warakulwit[2,3], Jumras Limtrakul[2,3], Tatsuya Tsukuda[4], Michael Probst[1,*]

[1] Institute of Ion Physics and Applied Physics, University of Innsbruck, Technikerstrasse 25, 6020 Innsbruck, Austria
[2] Chemistry Department, Center of Nanotechnology, and Center for Advanced Studies in Nanotechnology and Its Applications in Chemical, Food and Agricultural Industries, Kasetsart University, Bangkok 10900, Thailand
[3] NANOTEC Center of Excellence, National Nanotechnology Center, Kasetsart University, Bangkok 10900, Thailand
[4] Catalysis Research Center, Hokkaido University, Nishi 10, Kita 21, Sapporo 001-0021, Japan

**Abstract.** The thermal stabilization of thin gold nanowires with a diameter of about 2 nm by surfactants is investigated by means of classical molecular dynamics simulations. While the well-known melting point depression leads to a much lower melting of gold nanowires compared to bulk gold, coating the nanowires with surfactants can reverse this, given that the attractive interaction between surfactant molecules and gold atoms lies beyond a certain threshold. It is found that the melting process of coated nanowires is dominated by surface instability patterns, whereas the melting behaviour of gold nanowires in vacuum is dominated by the greater mobility of atoms with lower coordination numbers that are located at edges and corners. The suppression of the melting by surfactants is explained by the isotropic pressure acting on the gold surface (due to the attractive interaction) which successfully suppresses large-amplitude thermal motions of the gold atoms. (Note that this is a pre-peer-reviewed version which has been submitted to *Nanoscale*.)

## 1. Introduction

Interest in metal nanowires (NWs) stems from fields ranging from fundamental low-dimensional physics to technological applications in electronic, optoelectronic, nano-electromechanical, and nano-biotechnological systems [1-6]. Various experimental, theoretical and computational investigations with a variety of techniques have been conducted with the aim of revealing the special properties as well as possible technological applications of metal NWs. Examples are studies dealing with electronic metal-molecule-metal conjunctions [7, 8] and nano-electromechanical sensor systems [9]. For the purpose of nanoscale applications, a quantitative understanding of the structural and thermo-dynamic properties of such NWs is a prerequisite. Manipulating these tiny physical systems poses a challenge to many existing experimental techniques [10] and to complement experiments, a variety of computational studies have focused on thermal, mechanical, and electrical properties of metal nanosystems using mostly classical molecular dynamics (MD) simulation [11-27] and quantum chemical ab initio calculations [28-31].

---

[*] Correspondence to s.huber@uibk.ac.at and michael.probst@uibk.ac.at

One substantial difference between metal NWs and their bulk phase is the fact that the melting of NWs occurs at substantially lower temperatures than in the corresponding bulk material. This 'melting point depression' has been known for a long time from metal nanoparticles (NPs) [32-34]. Theoretical, experimental and simulation expertise has as well been adopted to explain the pre-melting of thin NWs [35]. The temperature dependence of structural properties of metal NWs and their melting behaviour have been investigated by various MD studies [13, 16, 18, 24, 25]. In these studies a strong dependence of the melting temperature and the melting behaviour on the size and the geometrical configuration of the metal NWs has been found. Similar dependences might influence other properties of metal NWs and also be related to technical applications. For example in [36] the stretching behaviour of gold NWs (AuNWs) subjected to external stresses was investigated. A solvent effect on the thinning of the wire was found only in the temperature region close to the melting point. Such findings can be important for technical applications, especially when the melting is strongly dependent on the size and configuration of the NW.

In this paper we use an atomistic model to investigate the melting process of thin AuNWs by means of MD simulations and some of the underlying physics. We also propose a model of the physical mechanism responsible for their thermal stabilization by coating them with surfactant molecules.

In section 2.1-3 we describe our computational model. In section 3.1 we discuss the results for the melting of a bare AuNWs in vacuum and the effects of additional initial defects such as dislocations. In section 3.2 we present our results concerning the effect of surfactants on the location of the melting point as well as the effect on how the melting proceeds. Section 4 summarizes our work.

## 2. Methodology

2.1 Molecular model

The interactions between gold atoms are modelled by the well-established quantum-corrected Sutton-Chen (Q-SC) many-body potential [37-39]. For this potential, the configuration energy of a system of atoms can be written as

$$U = \sum_i U_i = \sum_i \varepsilon \left[ \frac{1}{2} \sum_{j \neq i} V(R_{ij}) - c\sqrt{\rho_i} \right], \quad (1)$$

where $V(R_{ij})$ is a pair interaction function defined as

$$V(R_{ij}) = \left( \frac{a}{R_{ij}} \right)^n, \quad (2)$$

accounting for the repulsion between atoms $i$ and $j$, $\rho_i$ is a local density accounting for cohesion associated with atom $i$ defined by

$$\rho_i = \sum_{j \neq i} \Phi(R_{ij}) = \sum_{j \neq i} \left(\frac{a}{R_{ij}}\right)^m. \tag{3}$$

In (1)-(3), $R_{ij}$ is the distance between atoms $i$ and $j$; $a$ is a length parameter scaling all spacings such that $V$ and $\rho$ are dimensionless; $c$ is a dimensionless parameter scaling the attractive terms; $\varepsilon$ sets the overall energy scale; $n$ and $m$ are integer parameters such that $n > m$. We used for the Q-SC potential of gold the following values of the parameters: $n = 11$, $m = 8$, $\varepsilon = 7.8052$ meV, $c = 53.581$, and $a = 4.0651$ Å [25].

As a model surfactant n-aminopentane (AP: $NH_2$-$CH_2$-$CH_2$-$CH_2$-$CH_2$-$CH_3$) has been chosen. As a simplification, its amino group was assumed to interact like a methyl group within the AP subsystem. This avoids any contributions from hydrogen bonding and electrostatics in the simulation while the consequently poor description of the liquid AP phase is not important here. The total conformational energy, $V_{total}$, of AP can be written as

$$V_{total} = V_{intra} + V_{inter}, \tag{4}$$

where $V_{intra}$ contains in the usual way the bonded interactions arising from bond stretching, $V_{bond}$, bond bending, $V_{bend}$, and torsion, $V_{dihedral}$:

$$V_{intra} = V_{bond} + V_{bend} + V_{dihedral}, \tag{5}$$

with

$$V_{bond}(r_{ij}) = \sum_{ij} \frac{1}{2} k^r_{ij}(r_{ij} - r^{eq}_{ij})^2, \tag{6}$$

$$V_{bend}(\theta_{ijk}) = \sum_{ijk} \frac{1}{2} k^\theta_{ijk}(\theta_{ijk} - \theta^{eq}_{ijk})^2, \tag{7}$$

and

$$V_{dihedral}(\varphi_{ijkl}) = \sum_{ijkl} k^\varphi_{ijkl}[1 + \cos(m\varphi_{ijkl}) - \delta_{ijkl}], \tag{8}$$

where $r_{ij}$ is the bond distance between atoms $i$ and $j$, $r^{eq}_{ij}$ is the equilibrium bond length, $\theta_{ijk}$ is the angle defined by atoms $i$, $j$ and $k$, $\theta^{eq}_{ijk}$ is the equilibrium bond angle, $\varphi_{ijkl}$ is the dihedral angle defined by the four atoms $i$, $j$, $k$, and $l$, and $k^r_{ij}, k^\theta_{ijk}, k^\varphi_{ijkl}, \delta_{ijkl}$ and $m$ are parameters, which have been adopted from the CHARMM force field [40].

The non-bonded interaction in (4), $V_{inter}$ has been modelled in terms of 12-6 Lennard-Jones (LJ) potentials for all pair interactions between the various methyl, methylene and amino groups which are treated as united atom groups as well as the interaction of Au with $NH_2$, $CH_2$ and $CH_3$:

$$V_{\text{inter}} = \sum_{ij} 4\,\epsilon_{ij} \left[ \left(\frac{\sigma_{ij}}{r_{ij}}\right)^{12} - \left(\frac{\sigma_{ij}}{r_{ij}}\right)^{6} \right], \tag{9}$$

where $\sigma_{ij}, \epsilon_{ij}$ and $r_{ij}$ are the atomic or molecular group diameter, the depth of the potential well and the distance between atoms or molecular groups $i$ and $j$, respectively. For computational reasons the LJ potential is truncated and shifted smoothly to zero at a cut-off radius $r_c$ of about 10 Å. The values of the LJ parameters have again been adopted from the CHARMM force field for the AP constituents, and have been derived from the universal force field (UFF) [41] for the gold atoms. The parameters $\sigma$ for the cross interactions have then been calculated from the Lorentz-Berthelot combination rules. However, $\epsilon$ for the Au-NH$_2$ interaction was initially set to 4 kcal/mol which corresponds to about $6 \times k_B T_M / 2$, where $T_M$ is the melting temperature of the bare AuNW in vacuum and $k_B$ is Boltzmann's constant. Intra- and intermolecular parameters are summarized in table 1.

**Table 1.** Force field parameters [40, 41]

| bond | bond stretch $r_{ij}^{\text{eq}}$ (Å) | $k_{ij}^r$ (kcal/mol) | |
|---|---|---|---|
| CH$_3$(NH$_2$)-CH$_2$ | 225 | 1.54 | |
| CH$_2$-CH$_2$ | 225 | 1.52 | |

| angle | angle bend $\theta_{ijk}^{\text{eq}}$ | $k_{ijk}^\theta$ (kcal/mol) | |
|---|---|---|---|
| CH$_3$(NH$_2$)-CH$_2$-CH$_2$ | 110° | 45 | |
| CH$_2$-CH$_2$-CH$_2$ | 110° | 45 | |

| dihedral | torsion $\delta_{ijkl}$ | $k_{ijkl}^\varphi$ (kcal/mol) | $m$ |
|---|---|---|---|
| CH$_3$(NH$_2$)-CH$_2$-CH$_2$-CH$_2$ | 0° | 1.6 | 3 |
| CH$_2$-CH$_2$-CH$_2$-CH$_2$ | 0° | 1.6 | 3 |

| atom / group | non-bonded interaction (LJ) $\sigma$ (Å) | $\epsilon$ (kcal/mol) | |
|---|---|---|---|
| CH$_3$(NH$_2$) | 2.165 | 0.1811 | |
| CH$_2$ | 2.235 | 0.1142 | |
| Au | 2.934 | 0.039 | |

2.2 Simulation details

If not stated otherwise, the MD simulations have been performed in the canonical ensemble (NVT ensemble). For purposes of comparison and in order to exclude a certain pressure dependence of the results also several simulations have been performed in the isothermal-isobaric ensemble (NpT). Temperatures and/or pressures have been achieved by the Berendsen as well as the Nosé-Hoover thermo- and barostats [42-44].

In order to study the melting behaviour of a thin AuNW in vacuum as well as in contact with AP surfactants, different initial NW geometries have been prepared as discussed in the next subsection.

Their initial configurations have then been fully relaxed and have subsequently been subject to free evolution in a temperature range of 100 - 1500 K. Initially, temperature steps of 100 K were used to roughly determine the melting temperature and have then been adjusted to 20 K to for a more precise determination. The equations of motion have been integrated with the velocity Verlet algorithm with a variable time step in the order of $0.1 - 10$ fs. A simulated time of several hundred ps for each trajectory should ensure proper relaxation of the system at each temperature.

All simulations have been carried out with the DL_POLY 4 software [45].

2.3 Initial configurations

In order to investigate the melting process of a thin AuNW we first constructed a wire consisting of 200 layers consisting of 18 atoms each (3600 gold atoms in total) arranged in a perfect fcc lattice where the single layers correspond to the (001) surface. This model corresponds to a NW of about 40 nm in length and a cross section of about $1.2 \times 1.2$ nm$^2$. In addition to this model we constructed a NW with one dislocation to study the influence of this defect on the melting process (see the description of fig. 1 below). It should be mentioned again that our AuNW models are finite, i.e. the length of the simulation box is larger than the length of the simulated wire.

The exposition to surfactants was created by adding 3054 AP molecules, corresponding to a concentration of about 44g/l, and letting the system evolve freely at 600 K for several hundred ps. The resulting configuration exhibited a coverage of about 2/3. For computational purposes the 1406 AP molecules not coating the NW were then removed and only the coated NW has subsequently been used for the determination of the melting temperature for different values of the LJ-parameter $\epsilon$ describing the strength of the attractive interaction between gold atoms and amino groups. $\epsilon$ was varied from 1 to 7 kcal/mol in steps of 1 kcal/mol.

The three different initial configurations (perfect bare NW, bare NW with additional dislocation and coated NW) are shown in figure 1. The initialization procedure described above causes the AP amino groups to adsorb at the 4-fold hollow position on the gold (100/010/001) surfaces.

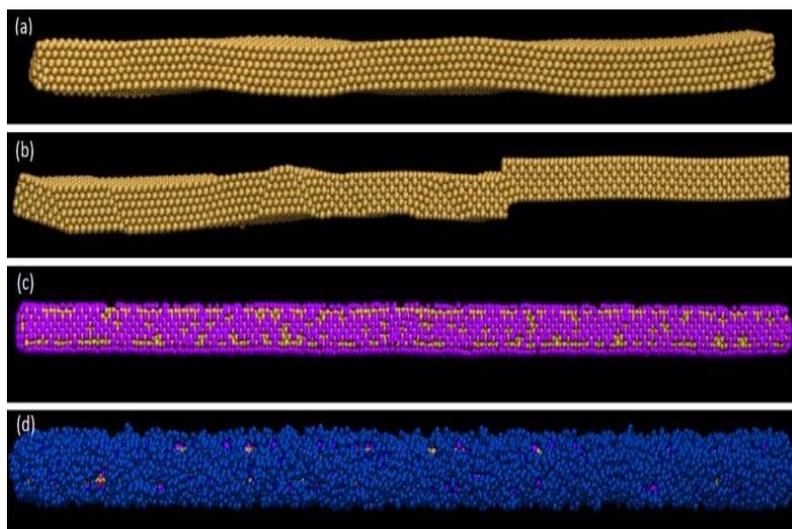

**Figure 1.** Three different initial configurations after relaxation at 100 K: (a) bare wire in vacuum, (b) bare wire with additional dislocation in vacuum, (c) coated NW (only the amino groups of AP molecules are shown), (d) coated NW (whole AP molecules are shown).

**3. Results**

3.1 Model verification

Before investigating the melting process of a bare NW in vacuum we checked the appropriateness and proper implementation of the Q-SC potential by investigating the melting of bulk gold. For ease of comparison with earlier publications [25] we simulated a bulk system consisting of 4000 gold atoms using the NpT-ensemble. The gold atoms were arranged in a perfect fcc structure in a cell subjected to cubic periodic boundary conditions. The phase transition from solid to liquid state around 1380 K can clearly be seen in the enthalpy-versus-temperature diagram of figure 2. Taking into account that some superheating does normally occur in surface-free perfect crystals [46, 47], the simulated melting point should be even closer to the experimental value of 1338 K for pure bulk Au [48]. The validity of the Q-SC potential for the melting behaviour of bulk gold implies also some confidence that it can also capture qualitative trends in finite systems.

This was further checked against available simulation data by simulating small gold nanoparticles (NPs) of 256 gold atoms arranged in a cuboid with an aspect ratio of 2:4:16 in a perfect fcc lattice. The melting temperature of this model was found at about 550 – 575 K (inset in figure 2), virtually the same as reported in previous studies [36, 49] with the - in principle more sophisticated - TB-SMA potential [50] for modelling the Au-Au interaction.

3.2 Melting of AuNWs in vacuum

After these checks we determined the melting temperature of the perfect bare NW described in section 2.3. The result of $T_M$=750 K is again best seen in the enthalpy-versus-temperature curve (figure 2). The enthalpy curve exhibits a drop at the temperature of the phase transition. This is an artefact due to the specific simulation details (size of the simulation cell, choice of the NpT ensemble, etc.). Since we are only interested in the temperature at which the melting of the NW begins, and since the qualitative reproduction of the melting point depression is not affected [35], no remedy of this behaviour was attempted. The simulated melting point depression resembles the finding in recent experiments [51] on gold NWs of similar size, although the instability of AuNWs observed there at much lower temperatures may not be associated with melting. Quantitative discrepancies are, however, most likely grounded in shortcomings of the analytic interaction potentials. They have been optimized to accurately describe bulk properties [39] and cannot be expected to handle systems outside the original realm in more than a qualitative fashion. This in itself is, of course, not surprising, given that, for example, not even a potential for water exists that spans larger regions of its phase diagram.

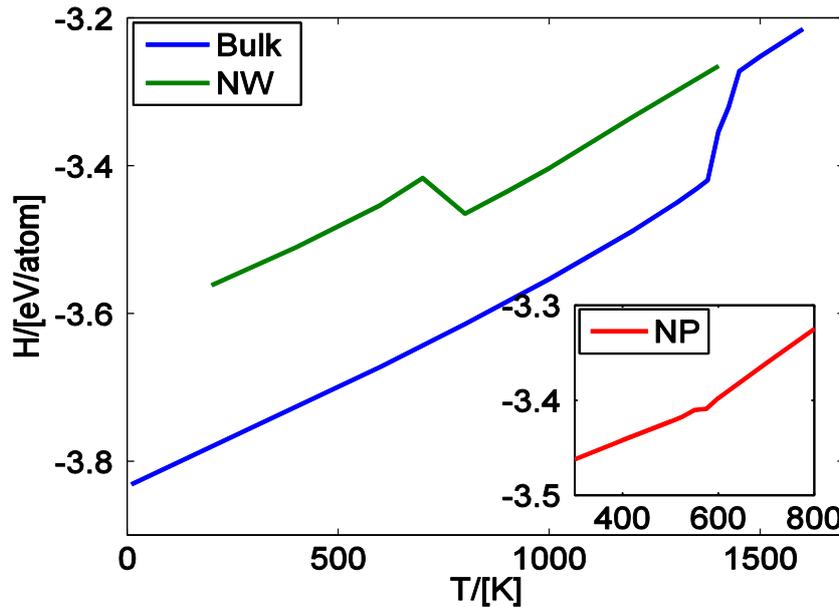

**Figure 2.** Calorimetric curves exhibiting the solid-liquid phase transition for bulk gold (blue), the bare AuNW (green) and AuNPs (inset, red).

Snapshots from the melting process of the bare NW in vacuum at 800 K are shown in figure 3. As stated in section 2.2, these simulations were carried out in the NVT ensemble. At first a thickening at both ends of the NW occurs. The ends move then towards each other until the system assumes a nugget-like and, finally, spherical shape. Concurrently, the middle region of the NW becomes unstable and wavy at the surface, thereby enhancing the melting procedure. However, the process that dominates the melting is the greater mobility of the gold atoms at the edges and corners with lower coordination numbers, i.e. on the two ends of the finite NW.

A simulation of a NW with an initial dislocation (figure 1b) at 750 K lends credibility to this. The effect of the defect at various temperatures is depicted in figure 4. The enhanced mobility of the gold atoms in the region of the initial dislocation, leads, after 30-50 ps at 750 K, to a 'shoulder' which develops into a kink. Also here the melting appears to be enhanced by instability patterns at the surface of the NW, but as in case of the perfect wire, the dominant process for the melting of the NWs in vacuum remains the greater mobility of atoms at edges and corners, i.e. the two ends as well as the defect region of the NW in this case. It should be mentioned that our simulations do not allow us to decide if there is a temperature region where solid and liquid phases coexist in a thermodynamically stable equilibrium.

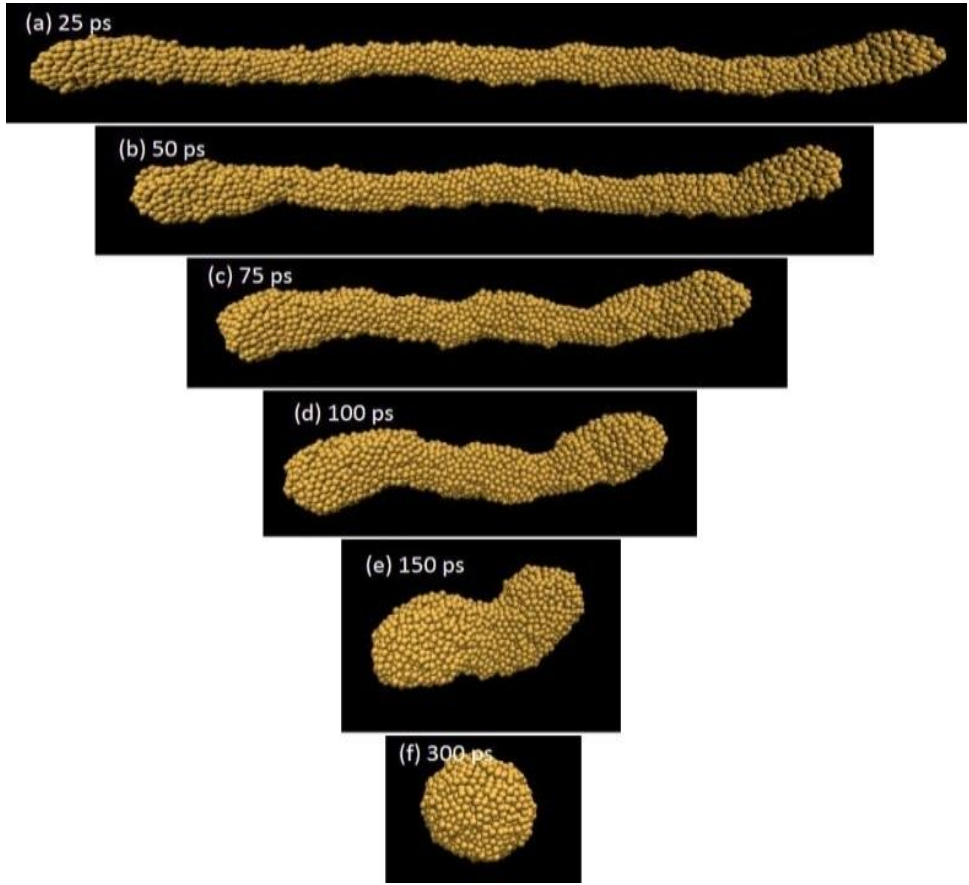

**Figure 3.** Melting process of the bare NW in vacuum (a) 25, (b) 50, (c) 75, (d) 100, (e) 150 and (f) 300 ps after initialization at 800 K.

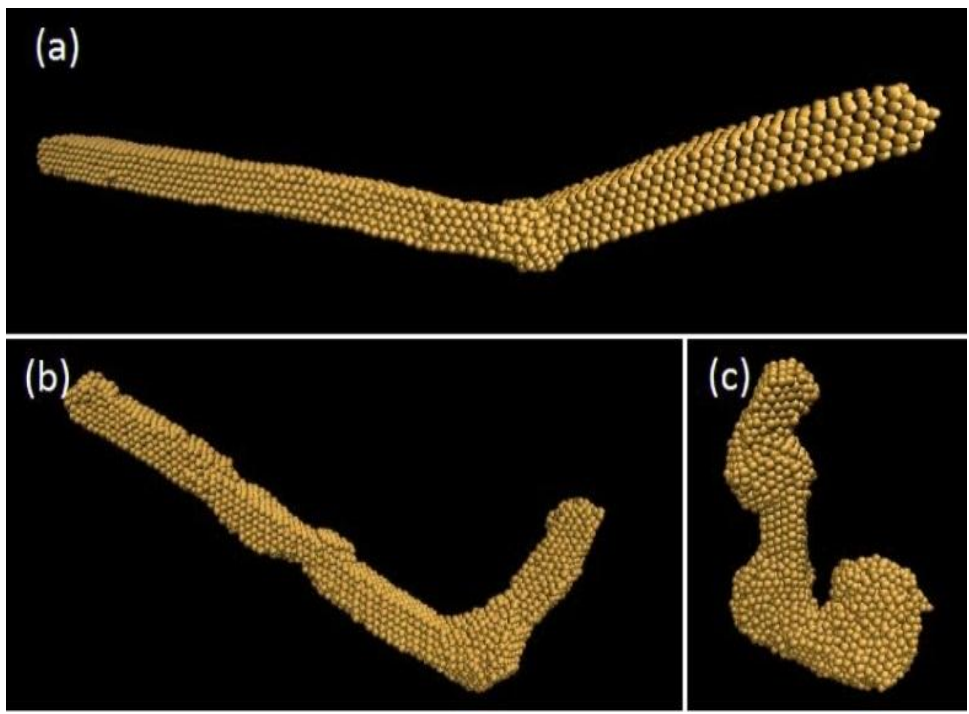

**Figure 4.** Effect of the additional initial dislocation at (a) 600, (b) 700 and (c) 750 K after several hundred ps. Although melting occurs only beyond 750 K the additional defect clearly affects the thermal behaviour of the NW.

3.3 Surfactant effects

The effect of adding surfactant molecules (AP) is best observed by plotting estimates (only a single time origin is taken into account) of the normalized root mean square displacement (NRMSD in Å per ps) as well as the diffusion coefficient (approximately corresponding to the derivative of the RMSD) of the gold atoms versus the temperature (figure 5). For an Au-$NH_2$ potential parameter $\epsilon < 4.5$ kcal/mol an abrupt change in the NRMSD, accompanied by the corresponding peak of the diffusion coefficient, can be observed in the temperature region around 750 K. However, for values of $\epsilon$ larger than 4.5 kcal/mol the jump in the NRMSD (peak in the diffusion coefficient) is shifted to substantially higher temperatures, indicating a corresponding increase of the melting temperature.

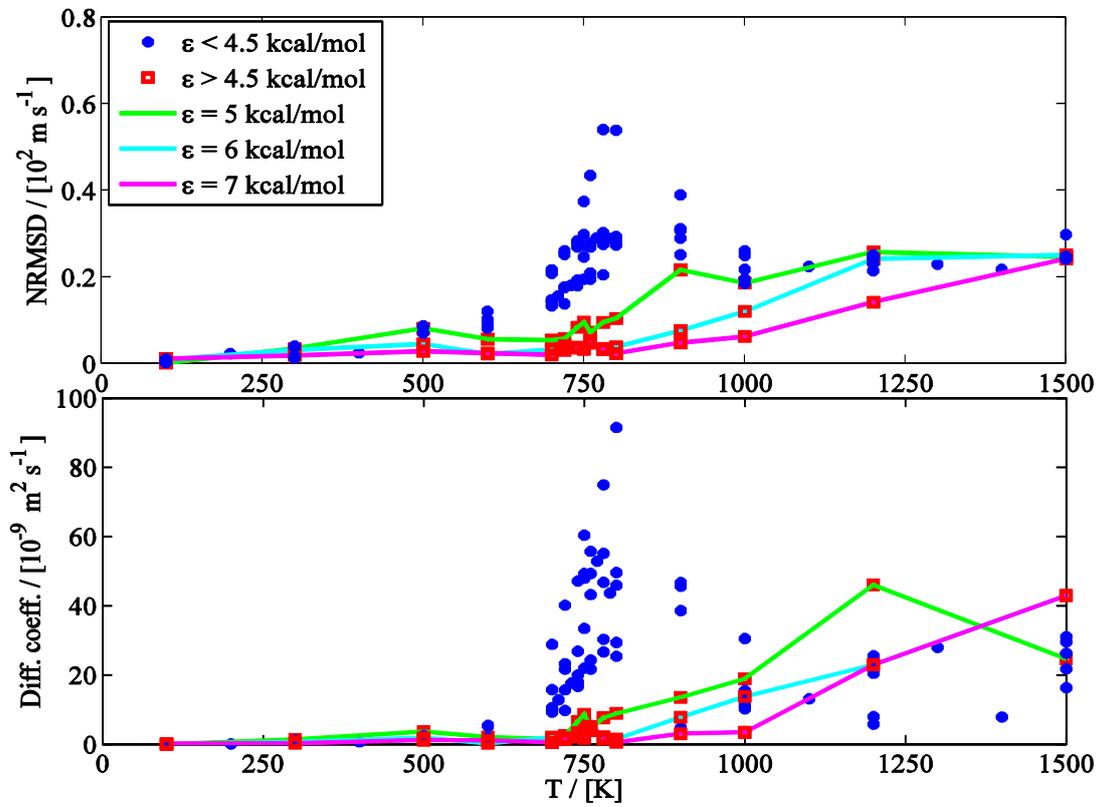

**Figure 5.** (Normalized) root mean square displacement (top) and diffusion coefficient (bottom) of the gold atoms at various temperatures and different values of $\epsilon$.

The threshold value for $\epsilon$ of about 4.5 kcal/mol equals approximately $6 \times k_B T_M/2$, where $T_M$ is the melting temperature of the bare AuNW in vacuum and $k_B$ is Boltzmann's constant. This corresponds nicely to the surfactant model which effectively gives rise to a behaviour of AP resembling a near-linear string with six degrees of freedom (three translational, two rotational and one vibrational). Equipartition of the energy leads then straightforwardly to the smallest necessary depth of the well of the LJ potential. The melting of the NW is suppressed until the AP molecules are no longer permanently absorbed and only a dynamical exchange of them on the surface remains. At temperatures $T < \epsilon/3k_B$ for $\epsilon > 4.5$ kcal/mol the isotropic pressure due to the attractive interaction between AP molecules and surface atoms suppresses the thermal motion of the gold atoms and thus, the melting of the NW.

This has also consequences for the structural evolution of a coated NW as can be seen from two further exemplary simulations at 800 and 1500 K with $\epsilon$ set to 6 kcal/mol. Figures 6 and 7 show snapshots from their evolution. Although the coated AuNW substantially changes its overall shape at 800 K, it remains elongated and no real melting occurs in contrast to the bare wire at the same temperature. While in the two-dimensional pictures of figure 6 it might appear that the wire gets shorter, a simulation movie shows that this effect is caused by the wiggling of the wire in space. At 1500 K the NW melts but differences compared to the bare NW can be observed: Whereas for the bare NW the greater mobility of the gold atoms at corners and edges has been the main melting mechanism, here dynamically evolving local imperfections in the coating of the NW with AP, which give subsequently rise to instability patterns at the NW surface, appear more important in bringing forward the melting procedure. A possible underlying physical mechanism in terms of a surface instability could be a Plateau-Rayleigh instability. However, this would have to be tested by more demanding simulations.

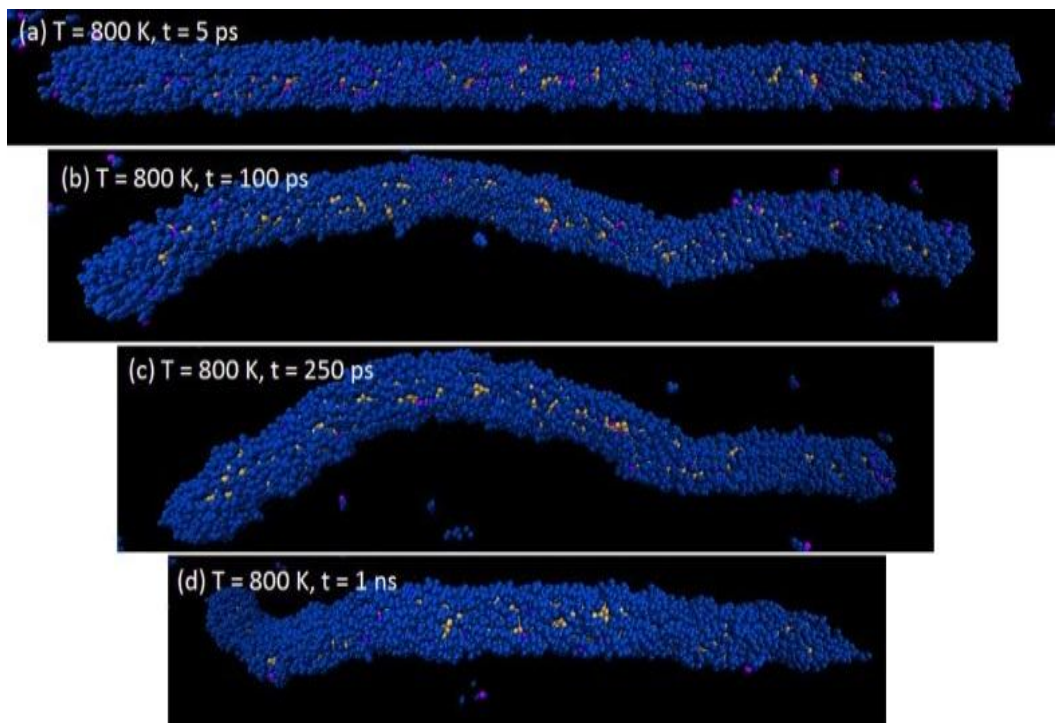

**Figure 6.** Surfactant-coated AuNW at 800 K at (a) 5, (b) 100, (c) 250 and (d) 1000 ps after initialization (AP molecules in blue).

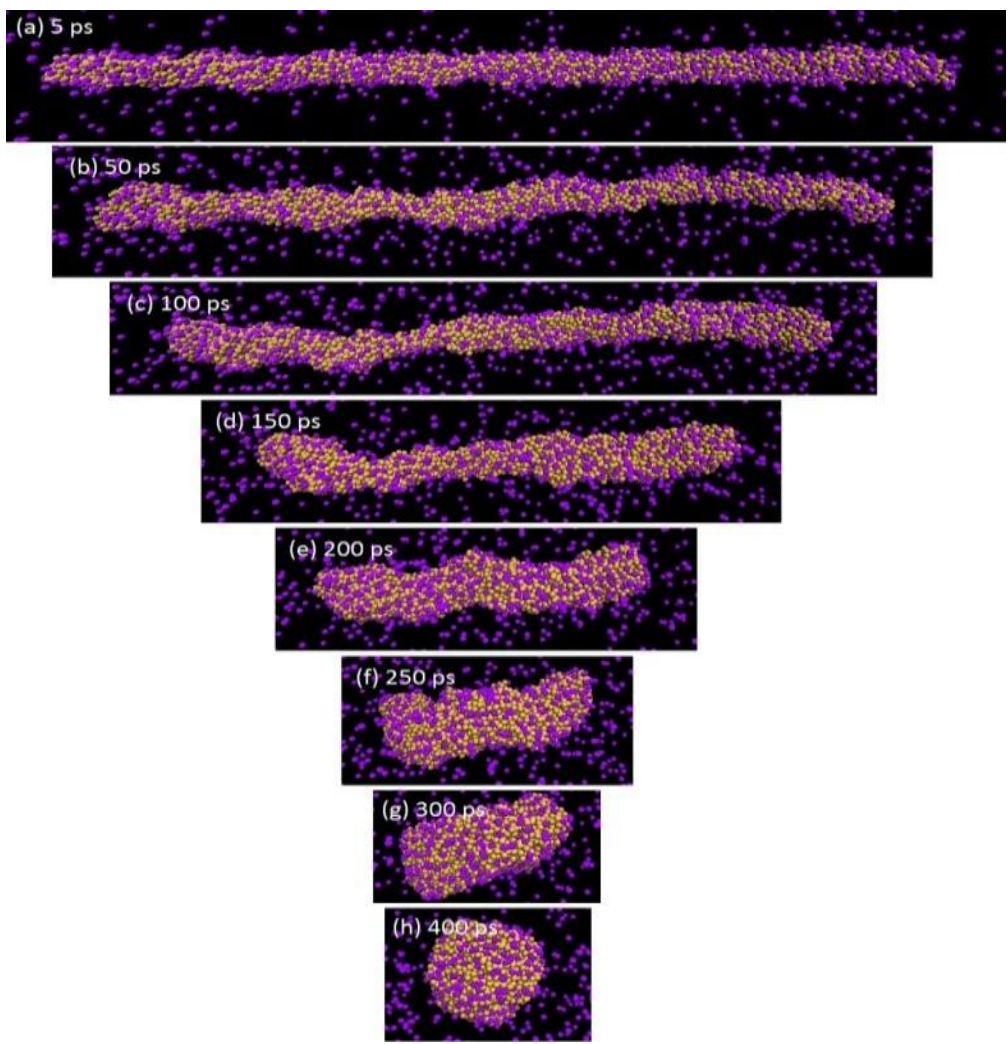

**Figure 7.** Coated NW at 1500 K and epsilon=6 kcal/mol (a) 5, (b) 50, (c) 100, (d) 150, (e) 200, (f) 250, (g) 300 and (h) 400 ps after initialization. Only the amino groups of AP molecules (violet) are shown for convenience.

## 4. Conclusion

We have investigated the melting behaviour of a thin AuNW in vacuum and in contact with a surfactant by means of a large number of MD simulations. For the bare NW the well-known melting point depression for metal nanosystems [35] leads to a lowering of $T_m$ from 1380 K for bulk gold to about 750 K in good agreement with other simulations. We found that the melting progress of the NW is dominated by the mobility of atoms located at the edges and corners of the finite NW rather than, for example, spontaneously formed defects. Initially present defects like a dislocation enhance the melting in their vicinity in the same way. A possible option to thermally stabilize thin AuNWs, i.e. to increase the depressed melting point, is the coating by surfactants. This has been investigated by submerging the NW into aminopentane as a model for larger surfactants. The attractive interaction between the amino groups of AP and the gold atoms is tuned to investigate the dependence of the melting temperature on that parameter. We observe an increase of the melting point depending on the strength of this attractive interaction. If the Lennard-Jones parameter $\epsilon$ exceeds $6 \times k_B T_M/2$, the melting point increases by 200-300 K. Inspection of the RMSD and diffusion coefficient of Au ensures that this is indeed the case and the system is not molten Au inside an AP micelle. Furthermore, we found a substantial difference between the melting kinetics of the bare NW and the NW in contact with the surfactant models: The latter is not dominated by the mobility of atoms at edges, corners and structural defects because the coating suppresses them. This in turn enhances the importance of spontaneously formed imperfections in the coating layer and related surface instabilities.

We conclude that the coating of metal NWs with surfactant molecules could be a method for the thermal stabilization of the NW in temperature regions where they would melt otherwise. Additionally, one might envisage that instead of classical surfactants, for certain applications, solid support materials e.g. carbon nanotubes [52-54] could be used to prevent premelting of AuNWs embedded into them and to provide mutual mechanical stabilisation.

## Acknowledgements

This work was supported by the Austrian Ministry of Science BMWF as part of the university infrastructure program of the scientific computing platform at LFU Innsbruck. Support from the DK+ on computational interdisciplinary modelling is gratefully acknowledged. C.W. and J.L. thank the National Science and Technology Development Agency (NSTDA Chair Professor and National Nanotechnology Center), the National Research University Project of Thailand (NRU) and the Thailand Research Fund for support. We thank Prof. Kersti Herrmannson for fruitful discussions and suggestions concerning our work.

## References

[1] Y. Kondo, K. Takayanagi, Science, 289 (2000) 606-608.
[2] Y. Kondo, K. Takayanagi, Physical Review Letters, 79 (1997) 3455-3458.
[3] B.H. Hong, S.C. Bae, C.-W. Lee, S. Jeong, K.S. Kim, Science, 294 (2001) 348-351.
[4] H.S. Park, W. Cai, H.D. Espinosa, H. Huang, MRS Bulletin, 34 (2009) 178-183.
[5] F. Patolsky, B.P. Timko, G. Zheng, C.M. Lieber, MRS Bulletin, 32 (2007) 142-149.
[6] C.M. Lieber, Z.L. Wang, MRS Bulletin, 32 (2007) 99-108.
[7] M.A. Reed, C. Zhou, C.J. Muller, T.P. Burgin, J.M. Tour, Science, 278 (1997) 252-254.
[8] A.I. Yanson, G.R. Bollinger, H.E. Van den Brom, N. Agrait, J.M. Van Ruitenbeck, Nature, 395 (1998) 783-785.
[9] H.G. Craighead, Science, 290 (2000) 1532-1535.
[10] Z.L. Wang, R.P. Gao, Z.W. Pan, Z.R. Dai, Advanced Engineering Materials, 3 (2001) 657-661.
[11] M.R. Sorensen, M. Brandbyge, K.W. Jacobsen, Physical Review B: Condensed Matter and Materials Physics, 57 (1998) 3283-3294.
[12] H. Ikeda, Y. Qi, T. Cagin, K. Samwer, W.L. Johnson, W.A. Goddard, III., Physical Review Letters, 82 (1999) 2900-2903.
[13] G. Bilalbegovic, Solid State Communications, 115 (2000) 73-76.
[14] Y.-H. Wen, Z.-Z. Zhu, R. Zhu, G.-F. Shao, Physica E: Low-Dimensional Systems & Nanostructures, 25 (2004) 47-54.
[15] L. Hui, B.L. Wang, J.L. Wang, G.H. Wang, Journal of Chemical Physics, 121 (2004) 8990-8996.
[16] L. Hui, B.L. Wang, J.L. Wang, G.H. Wang, Journal of Chemical Physics, 120 (2004) 3431-3438.
[17] H. Liang, M. Upmanyu, H. Huang, Physical Review B: Condensed Matter and Materials Physics, 71 (2005) 241403/1-241403/4.
[18] L. Miao, V.R. Bhetanabotla, B. Joseph, Physical Review B: Condensed Matter and Materials Physics, 72 (2005) 134109/1-134109/12.

[19] S.J.A. Koh, H.P. Lee, C. Lu, Q.H. Cheng, Physical Review B: Condensed Matter and Materials Physics, 72 (2005) 085414/1-085414/11.
[20] H.S. Park, J.A. Zimmerman, Physical Review B: Condensed Matter and Materials Physics, 72 (2005) 054106/1-054106/9.
[21] H.S. Park, C. Ji, Acta Materialia, 54 (2006) 2645-2654.
[22] S.J.A. Koh, H.P. Lee, Nanotechnology, 17 (2006) 3451-3467.
[23] Y.-H. Wen, Z.-Z. Zhu, R. Zhu, Computational Materials Science, 41 (2008) 553-560.
[24] Y.-H. Wen, Y. Zhang, J.-C. Zheng, Z.-Z. Zhu, S.-G. Sun, Journal of Physical Chemistry C, 113 (2009) 20611-20617.
[25] Y. Zhang, Y.-H. Wen, J.-C. Zheng, Z.-Z. Zhu, Physics Letters A, 373 (2009) 3454-3458.
[26] Y. Zhang, Y.-H. Wen, Z.-Z. Zhu, S.-G. Sun, Journal of Physical Chemistry C, 114 (2010) 18841-18846.
[27] L. Wu, Y. Zhang, Y.-H. Wen, Z.-Z. Zhu, S.-G. Sun, Chemical Physics Letters, 502 (2011) 207-210.
[28] E.Z. Da Silva, F.D. Novaes, A.J.R. Da Silva, A. Fazzio, Physical Review B: Condensed Matter and Materials Physics, 69 (2004) 115411/1-115411/11.
[29] T. Pawluk, Y. Hirata, L. Wang, Journal of Physical Chemistry B, 109 (2005) 20817-20823.
[30] M. Palummo, S. Ossicini, R. Del Sole, Physica Status Solidi B: Basic Solid State Physics, 247 (2010) 2089-2095.
[31] A. Srivastava, N. Tyagi, R.K. Singh, Materials Chemistry and Physics, 127 (2011) 489-494.
[32] P. Pawlow, Zeitschrift fuer Physikalische Chemie, Stoechiometrie und Verwandtschaftslehre, 65 (1909) 1-35.
[33] J.P. Borel, Surface Science, 106 (1981) 1-9.
[34] T.P. Martin, Physics Reports, 273 (1996).
[35] O. Guelseren, F. Ercolessi, E. Tosatti, Physical Review B: Condensed Matter and Materials Physics, 51 (1994) 7377-7380.
[36] Q. Pu, Y. Leng, X. Zhao, P.T. Cummings, Nanotechnology, 18 (2007).
[37] A.P. Sutton, J. Chen, Philosophical Magazine Letters, 61 (1990).
[38] H. Rafii-Tabar, A.P. Sutton, Philosophical Magazine Letters, 63 (1991).
[39] B.D. Todd, R.M. Lynden-Bell, Surface Science, 281 (1993).
[40] P.v.R. Schleyer, e. al., The Encyclopedia of Computational Chemistry, in, John Wiley & Sons, Chichester, 1998, pp. 271-277.
[41] A.K. Rappe, C.J. Casewit, K.S. Colwell, W.A. Goddard, W.M. Skiff, J. Am. Chem. Soc., 114 (1992).
[42] W.G. Hoover, Physical Review A, 31 (1985).
[43] S. Melchionna, G. Ciccotti, B.L. Holian, Molecular Physics, 78 (1993) 533-544.
[44] H.J.C. Berendsen, J.P.M. Postma, W.F. Van Gunsteren, A. DiNola, J.R. Haak, Journal of Chemical Physics, 81 (1984) 3684-3690.
[45] I.T. Todorov, W. Smith, K. Trachenko, M.T. Dove, J. Mater. Chem., 16 (2006) 1911-1918.
[46] A. Boutin, B. Rousseau, A.H. Fuchs, Europhysics Letters, 18 (1992) 245-250.
[47] Z.H. Jin, P. Gumbsch, K. Lu, E. Ma, Physical Review Letters, 87 (2001) 055703/1-055703/4.
[48] C. Kittel, Introduction to Solid State Physics, Wiley, New York, 1996.
[49] Q. Pu, Y. Leng, L. Tsetseris, H.S. Park, S.T. Pantelides, P.T. Cummings, The Journal of Chemical Physics, 126 (2007) 144707/1-144707/6.
[50] F. Cleri, V. Rosato, Physical Review B: Condensed Matter and Materials Physics, 48 (1993) 22-33.
[51] C. Warakulwit et al., to be published.
[52] M.-F. Ng, J. Zheng, P. Wu, Journal of Physical Chemistry C, 114 (2010) 8542-8545.
[53] C.-H. Lee, C.-K. Yang, Journal of Physical Chemistry C, 115 (2011) 10524-10530.
[54] S.-F. Wang, Y. Zhang, L.-Y. Chen, J.-M. Zhang, K.-W. Xu, Physica Status Solidi A: Applications and Materials Science, 208 (2011) 97-103.